\documentclass[acmtog,nonacm]{acmart}
\acmSubmissionID{1234}

\usepackage{booktabs} 

\usepackage[ruled]{algorithm2e} 

\SetAlFnt{\small}
\SetAlCapFnt{\small}
\SetAlCapNameFnt{\small}
\SetAlCapHSkip{0pt}

\acmJournal{TOG}

\usepackage{xcolor}
\usepackage{listings}
\usepackage{physics}

\definecolor{mGreen}{rgb}{0,0.6,0}
\definecolor{mGray}{rgb}{0.5,0.5,0.5}
\definecolor{mPurple}{rgb}{0.58,0,0.82}

\lstdefinestyle{CStyle}{
    commentstyle=\color{mGreen},
    keywordstyle=\color{magenta},
    numberstyle=\tiny\color{mGray},
    stringstyle=\color{mPurple},
    basicstyle=\footnotesize,
    breakatwhitespace=false,         
    breaklines=true,                 
    captionpos=b,                    
    keepspaces=true,                 
    numbers=left,                    
    numbersep=5pt,                  
    showspaces=false,                
    showstringspaces=false,
    showtabs=false,                  
    tabsize=2,
    language=C
}

\begin{document}
\title{General purpose graphical rendering on quantum devices with composable function systems}

\author{James Schloss}
\authornote{These authors contributed equally to the paper.}
\orcid{0000-0002-3243-8918}
\affiliation{%
  \institution{MIT}
  \country{USA}}
\email{jars@mit.edu}

\author{Ayaka Usui}
\authornotemark[1]
\orcid{0000-0002-2326-3917}
\affiliation{%
  \institution{Universitat Aut\`{o}noma de Barcelona}
  \country{Spain}}
\email{ayaka.usui@uab.cat}

\begin{abstract}
The controlled creation of specific quantum states is a highly challenging field of research that is also in high demand with applications in various quantum technologies.
Due to its difficulty, it has been historically impossible to use quantum states as a rendering target for complex scenes and visualizations, especially in the Noisy Intermediate-Scale Quantum (NISQ) era where operations are noisy and only a few qubits are available.
Even so, in this paper we propose a new, quantum compatible method for general purpose rendering by extending Composable Function Systems (CFSs) to quantum architectures.
We discuss limitations in the classical implementation of CFSs and the physical maps necessary for adapting the classical method to quantum by showing the generation of object primitives and transformations of such objects, including duplications and smears, some of which are topologically non-trivial.
Our results reveal the possibility of a speed-up for this, specific method on quantum hardware and our method has been used to create the first video rendered on quantum architectures.

\end{abstract}

%
%

\ccsdesc[500]{Computing methodologies~Rendering}
\ccsdesc[500]{Hardware~Quantum computation}

%
%

\keywords{Iterated Function Systems, quantum, rendering}

\begin{teaserfigure}
\includegraphics[width=\textwidth]{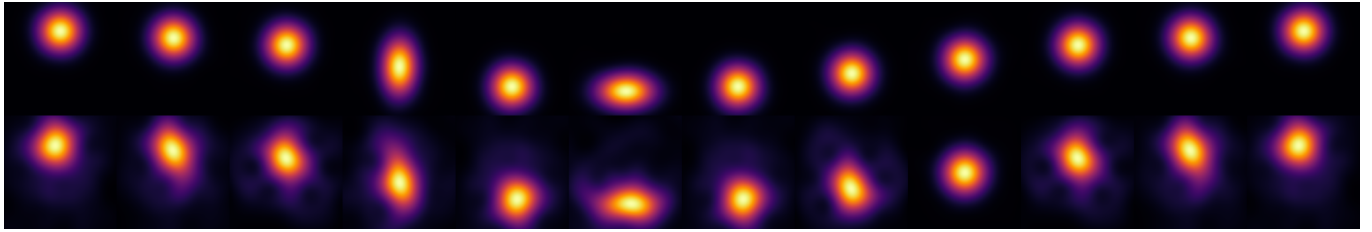}
\caption{Frames from the first video rendered on quantum hardware (IBM's Kingston system) with Composable Function Systems. Top: emulated results with 4 qubits per mode. Bottom: experimental results.}
\label{fig:video}
\end{teaserfigure}

\maketitle 

\section{Introduction}
\label{sec:intro}

General purpose rendering techniques allow for the controlled creation of various graphics and serve as a target for physical modeling.
Meshes, for example, act as both a method for artists to render $n$-dimensional scenes and as useful tools for various physical simulations.
Many alternative rendering methods exist, including raytracing, raymarching, splatting, and function systems, all of which have various advantages and have been used for complementary physical simulations.
Even so, no such general purpose rendering framework has been proposed for quantum architectures.
This paper details the creation of the first general-purpose rendering method for quantum systems as well as the creation of the first video rendered on quantum architectures by using IBM's Kingston system (Figure~\ref{fig:video} and supplementary video~\cite{suppvid}).
Our technique is inspired by a classical function system method known as Composable Function Systems (CFSs)~\cite{schloss2026cfs}, which do not necessitate the storage of mesh points or point clouds to memory, and our corresponding technique is thus possible to use in the Noisy Intermediate-Scale Quantum (NISQ) era, where operations are noisy and only a few qubits are available.
Our method is capable of drawing complex scenes shown in Figure~\ref{fig:emuatom}, and our results reveal a potential speedup for this specific rendering technique on quantum hardware.
This signals a meaningful advancement in the capabilities of quantum computation for every day use and could allow for the creation of (for example) quantum desktop environments or other interactive utilities.
We hope that this method will serve as a starting point for both artistic applications and physical simulations on quantum architectures.

Function systems, in general, are an interesting, yet under-explored method for graphical rendering with decades of rich history for various applications~\cite{barnsley1985iterated, barnsley2011chaos, fisher1994fractal, elliott2003functional}.
These methods are advantageous in that they do not require reservoirs of memory to generate objects.
Meshes, for example, require vertices.
Clouds require points.
Rather than storing these structures in global memory, function systems allow for the algorithmic generation of such structures on-the-fly.
It is important to note that methods with similar goals exist in the literature, such as with raymarching~\cite{slusallek2005introduction}, task-graph generation of mesh shaders~\cite{kuth2024real}, and closed-form implicit surfaces~\cite{keeter2020massively}; however, these methods either require complex physical simulations (ray marching, implicit surfaces) or expand into a memory reservoir to be used later in the rendering pipeline (mesh shaders).
These restrictions make them ill-suited for NISQ-era quantum architectures.
CFSs are unique in that they are intended to operate on various object primitives, such as those created with Iterated Function Systems (IFSs)~\cite{ghosh2022iterated}, for which quantum analogues already exist with Quantum Iterated Function Systems (QIFS)~\cite{lozinski2003quantum, jadczyk2004quantum}.
Moreover, because CFSs are compatible with quantum architectures, improvements to the classical algorithm may also be reflected in the quantum formulation.
All code for the classical method can be found in ~\cite{quibbledocs} and the quantum method in ~\cite{qifs}.

This paper is organized in the following way.
In Section~\ref{sec:related}, we discuss related methods.
In Section~\ref{sec:cfs}, we discuss the CFS method and limitations on classical hardware.
In Section~\ref{sec:quantum}, we move on the quantum version and discuss the basic formulation of the method, including visualization (Section~\ref{sec:husimi}), the generation of different object primitives (Section~\ref{sec:primitives}), affine transformations (Section~\ref{sec:affine}),  general transformations (Section~\ref{sec:gp}), and a fully integrated example on quantum hardware (Section~\ref{sec:example}).
In Section~\ref{sec:limits}, we discuss limitations to this method and areas of future work.
Finally, we conclude in Section~\ref{sec:conclusion}.

\section{Related work}
\label{sec:related}

In terms of quantum visualizations, there are no such methods in the literature for general purpose rendering on quantum hardware; however, there are related methods that are worth mentioning.
The controlled creation of quantum states has been a common research objective for many related areas of physical modeling for decades~\cite{weinacht1999controlling} and there are many examples of such in the literature.
For example ~\cite{bao2024creating} creates two-component Schr\"odinger cat (Greenberger-Horne-Zeilinger) states, ~\cite{schloss2020controlled} creates specific superfluid states, and ~\cite{siegl2022controlled} creates skyrmions.
It should be noted that many of these methods require specific experimental set-ups and are not suitable for NISQ-era Quantum Processing Units (QPUs).
A more closely related method is spin squeezing, which is used to engineer highly non-classical states on the Bloch sphere or in phase space, e.g.~\cite{kitagawa1993squeezed,ma2011quantum,pezze2018quantum,montenegro2025metrology}.
In addition, the concept of Quantum Iterated Function Systems (QIFSs) has been discussed previously by a few different groups~\cite{lozinski2003quantum, jadczyk2004quantum}.
There are also a number of papers using a specific IFS, the baker's map, to study various phenomenon in quantum systems~\cite{scott2003bakers, balazs1987bakers, pakonski1999bakers,lozinski2002irreversible,balazs1989the}.
Similar to the classical CFS paper~\cite{schloss2026cfs}, this work allows for the reformulation of QIFSs for use as object primitives and extends them into a general-purpose framework.

There are also a few methods related to quantum image processing~\cite{wang2022imagereview}, including the formulation of various image representation formats, quantum image compression ~\cite{deb2024imagecompress, roncallo2023jpeg, latorre2005image}, and watermarking~\cite{dhar2024watermark}.
Our representation differs from these methods in its construction and method for output in order to minimize the number of qubits.
Importantly, many of the aforementioned studies use the QPU as an accelerator for image processing of classical images and therefore need to be interoperable with existing image formats.
In this study, we instead focus on CFSs, which allow for image representations as function systems and do not need to store images as (for example) arrays with red, green, blue, and alpha channels or other common formats.
However, as more qubits become commercially available, it may be beneficial to use existing qubit image representation methods for faster visualization. Moreover, we will comment on faster visualization for our approach in Section~\ref{sec:limits}.

It is important to note that the classical CFS paper~\cite{schloss2026cfs} introduces a metaprogramming framework for rendering CFSs, and this method similarly introduces straightforward mathematical concepts for rendering specific images.
For this reason, it is important to highlight other software ecosystems that are attempting to create high-level abstractions for quantum architectures.
In particular, there are several efforts to create a streamlined quantum compiler for general-purpose computation, such as with PennyLane~\cite{bergholm2018pennylane, hopf2026integrating}, Qiskit~\cite{javadi2024quantum}, and OpenCLQPU~\cite{vazquez2024qpu}, and QLLVM~\cite{zhu2026qllvm}.
The most similar software framework to this study is bosonic qiskit~\cite{stavenger2022c2qa}, which allows for the creation of Fock states on qubit systems for use in hybrid bosonic-digital systems.
For this study, we have chosen to work primarily with Qiskit, itself, to run our generated circuits on hardware.
Though we intend to create a more streamlined software experience in future work, at the current time any of these software frameworks can be used to implement quantum CFSs.

\begin{figure}
    \includegraphics[width=0.4\textwidth]{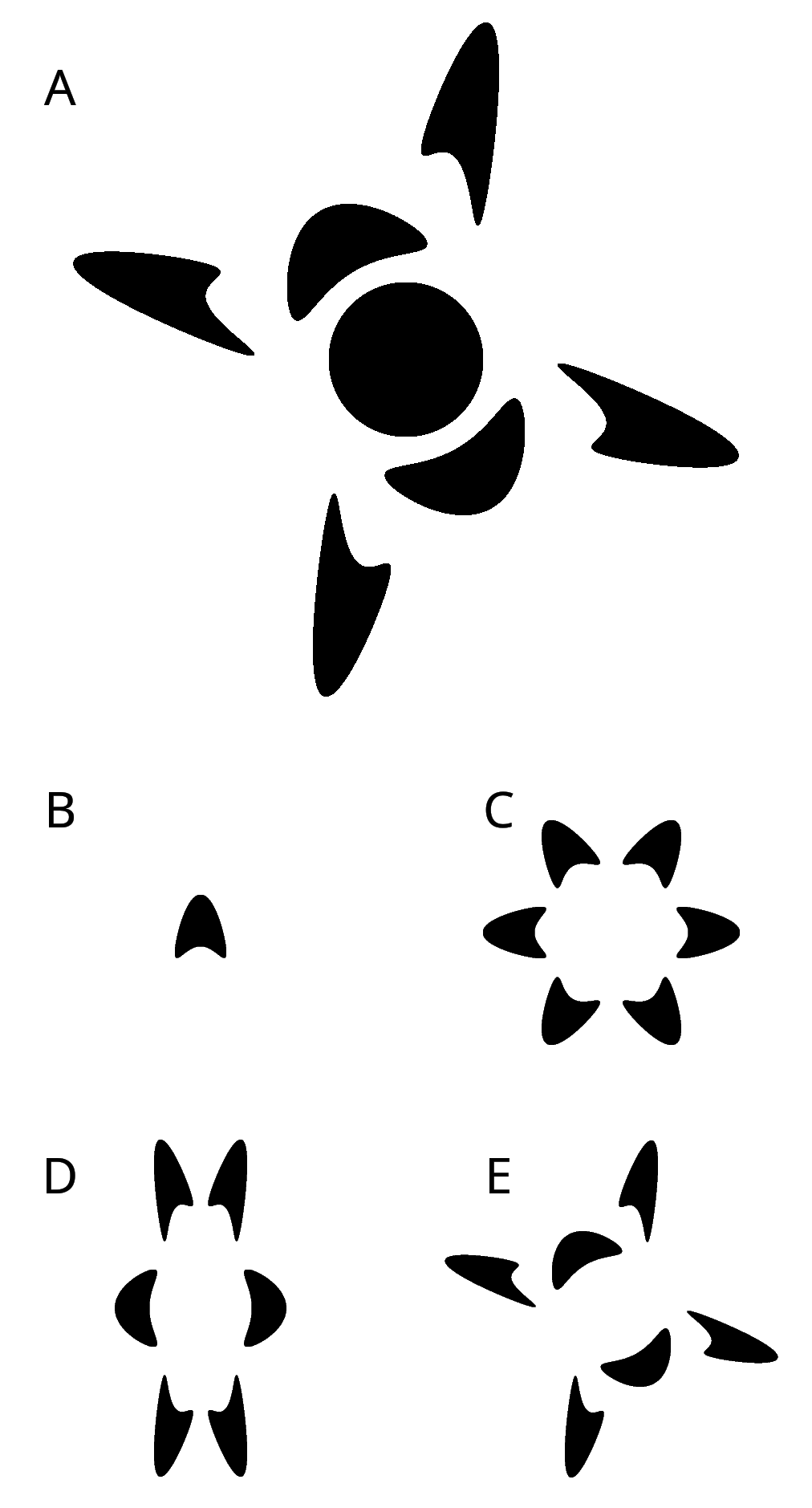}
    \caption{A frame from a dynamic ``atom'' visualization. All transformations are described in text.}
    \label{fig:classical}
\end{figure}

\section{Classical Composable Function Systems and limitations}
\label{sec:cfs}

Classical CFS frameworks are typically composed of two separate function sets for (1) the creation of the object primitives and (2) the transformation of that primitive into various shapes.
The primitives in CFS frameworks can be thought of as point clouds that do not need to be stored in a large memory reservoir because transformations are performed during the generation of the cloud, itself.
This is distinctly different than traditional mesh-based rendering strategies like mesh and geometry shaders in that CFSs do not necessitate the storing of intermediary data (vertices for meshes) to memory.
It is for this reason that CFSs are compatible with NISQ-era quantum devices.
To highlight the typical CFS workflow, let us detail the creation of the atom-like object in Figure~\ref{fig:classical}.
This figure will be later ported to quantum CFSs in Section~\ref{sec:example}.

For CFSs, the initial primitive may be generated in any way; however, let us consider the IFS for a circle:
\begin{enumerate}
    \item Transformation to polar coordinates 
          \begin{align}
              r &= \sqrt{x^2 + y^2} \\
              \theta &= \atan(y  / x)
          \end{align}
    \item Creation of square of variable density
          \begin{align}
              \theta_2 &= \pi (r^2 + f_{id}) \\
              r_2 &= \sqrt{\frac{\theta}{2\pi}}
          \end{align}
    \item Transformation back to Cartesian coordinates
          \begin{align}
              x_2 &= r^2\cos(\theta_2) \\
              y_2 &= r^2\sin(\theta_2)
          \end{align}
\end{enumerate}
Here, $x$ and $y$ are the positions of the point, $r$ is the distance from the origin, and $\theta$ is its corresponding angle.
The values $x_2$, $y_2$, $r_2$, and $\theta_2$ represent the values of the next point in the iteration and $f_{id}$ represents either $0$ or $1$ depending on which function is chosen.
This function system can be interpreted as a map from the original circle on to either the top or bottom half of itself.
A more complete description of this IFS can be found in ~\cite{schloss2026cfs}.
This IFS is not unique there are different circle constructions for different purposes; however, all constructions for circles require either non-affine or expansive maps, which are atypical for IFSs~\cite{lesniak2025highly}.

From this primitive, we can now introduce several transforms to create various objects, but we should note some nuances to this approach in Appendix~\ref{app:classysquare} and Figure~\ref{fig:square}.
For the construction in Figure~\ref{fig:classical}, we require several transforms for the outer electrons (lettering corresponds to those in the figure):

\begin{description}
    \item[B] A warp: $y = y + x^2$.
    \item[C] A displacement and rotation set: $y = y -2r$ as well as rotations about the origin to create the electrons. A global rotation is then added depending on the frame for movement seen in the supplementary video~\cite{suppvid}
    \item[D] A stretch or squeeze such that $x_2 = x/s$ and $y_2 = sy$ where $s$ is some stretching factor.
    \item[E] Additional rotations are performed for pairs of electrons to create orbits
\end{description}

Once all these operations are performed, we simply add back the unit circle to create a nucleus-like object for Figure~\ref{fig:classical}(A).

It is important to note that there are several limitations of CFSs on classical hardware, many of which are related to the initial drawing of the object primitive, itself.
Though no specific method of generating the primitive is preferred by CFS frameworks, \cite{schloss2026cfs} primarily relies on IFSs solved with the chaos game~\cite{barnsley2011chaos}, which requires a random function to be chosen each step.
After a few thousand iterations, the final object (known as the attractor) is visible on screen.
Such a method is ill-suited for massively parallel hardware (such as Graphics Processing Units (GPUs)) in the following ways:

\begin{enumerate}
    \item Function pointers are difficult to implement on GPUs~\cite{zhang2021judging} and there are few available programming interfaces that allow for their use without restrictions.
    \item Composing function systems together can be tricky because compilers struggle to properly inline transformation functions. This means that some form of metaprogramming is often necessary for CFSs to be used in practice. To overcome this, Quibble, the interface proposed in ~\cite{schloss2026cfs, quibbledocs}, opts for a straightforward approach that directly transpiles to OpenCL kernels before compiling at runtime.
    \item Random number generation per thread is costly.
    \item Allowing each thread to choose a function at random will (almost assuredly) lead to warp divergence and cut the performance by a factor equivalent to the number of functions used in the IFS for the primitive's construction.
\end{enumerate}

In the following section, we will discuss how these limitations can be properly addressed on quantum hardware.

\section{Quantum formulation of composable function systems}
\label{sec:quantum}

In this section, we will sketch out the necessary rendering pipeline for our quantum formulation of CFSs and differentiate it from the classical version.
The largest difference between the two methods is conceptual.
CFSs on classical hardware involve manipulating points and either splatting or histogramming them to screen one step at a time.
On quantum hardware, we instead manipulate the wavefunction, itself, to draw a probability distribution on a particle's phase and position after all computation is completed.
This means that the compilation pipeline for quantum is much more straightforward and does not require function pointers or metaprogramming as in the classical case (which addresses points 1 and 2 from Section~\ref{sec:limits}).
Instead, we can simply append the necessary transformation functions to the end of our primitive computation method.
Moreover, randomness is an inherent property of quantum measurement (addressing point 3), so the random selection of functions for the chaos game is more straightforward, especially because there are no concepts of warps to diverge (addressing point 4).
For these reasons, CFSs can be considered quantum compatible,
and there is potential of improvement over the classical version due to inherent properties of quantum processors.

\subsection{Visualization of quantum states with Husimi function}
\label{sec:husimi}

The general workflow for quantum CFS implementations is to first directly manipulate the wavefunction of our quantum state and then output its probability distribution as an image.
Though there are many representations of images on qubit-based systems~\cite{wang2022imagereview}, in this study, we have used the Husimi function directly as it is both straightforward to implement and does not require additional storage qubits~\cite{husimi1940}.
This method can be thought of as the Wigner function~\cite{wigner1932quantum} convolved with a Gaussian blur and thus has useful anti-aliasing effects.
This means that visualizations shown in this work will be notably blurrier than their classical counterpart; however, this we can choose non-Gaussian functions to mitigate this effect in future work.

Simply put, the Husimi function is a fidelity measurement with coherent states at specific locations along phase space (position and momentum plane).
More specifically, we denote $\rho$ as the state of visual interest and compute the following for different coherent states:
\begin{equation}
    H_\rho(p,q) 
    = 
    \mel{q,p}{\rho}{q,p},
\end{equation}
where $\ket{q,p}$ is a coherent state with $q$ position and $p$ momentum and is defined in Fock basis $\{\ket{n}_{\text{Fock}}\}$ as
\begin{equation}
    \ket{q,p}
    =
    e^{-\frac{q^2+p^2}{2}}
    \sum_{n=0}^{\infty}
    \frac{\left(q+ip\right)^n}{\sqrt{n!}}
    \ket{n}_{\text{Fock}}
    ,
\end{equation}
which is expressed in an infinite-dimensional system but can be also approximated in a truncated finite system.
The coherent state $\ket{q,p}$ can be considered the classical point $(q,p)$, and each location of $H_{\rho}(p,q) $ constitutes a single pixel in the output image. The movement between each location in the plane is performed classically, and we can therefore change frame of reference easily.
It is important to remind the reader that fidelity will be one only if the two input states are the same beside the global phase.
In this way, the Husimi function is constructed with measurement of the fidelity by changing $q,p$, and the maximum ranges of $q,p$ the user can take are fixed by the system size of the truncated system.

The Husimi function and Fock basis are often used in (but not limited to) quantum optics, and some of the operators introduced in this work have been well-studied and implemented in the field of continuous variables, e.g.~\cite{dodonov2002nonclassical}.
Nonetheless, there is no problem in using the Fock basis on qubit-based quantum architectures or hardware due to its universality.
Moreover, photonic computing with continuous variables is still being actively developed, e.g.~\cite{madsen2022quantum,aghaee2025scaling}.
It would be interesting to see how our approach can be implemented in photonic hardware when it is ready.
For this study, we will limit ourselves to qubit-based architectures, particularly IBM hardware.




\subsection{Generating object primitives on quantum}
\label{sec:primitives}

Primitives are objects intended to be manipulated later in the graphical rendering pipeline by (in the case of CFSs) arbitrary methods imposed by the user. In the classical method, primitives can be any designed shape, but circles, squares, and triangles allow for the majority of visualizations necessary for general-purpose rendering. In this section, we will detail the construction of both circles and squares in quantum systems.
For the construction of circles, we take advantage of the fact that Fock states are represented in phase space as a Gaussian or a ring, both of which may also be treated as primitives for this method.
The construction of squares, in particular, is inspired by \cite{lozinski2003quantum,schloss2026cfs}, and iteratively performed.
We will discuss both separately.



\subsubsection{Circles} \label{sec:circles_quantum}
\begin{figure}
    \includegraphics[width=0.4\textwidth]{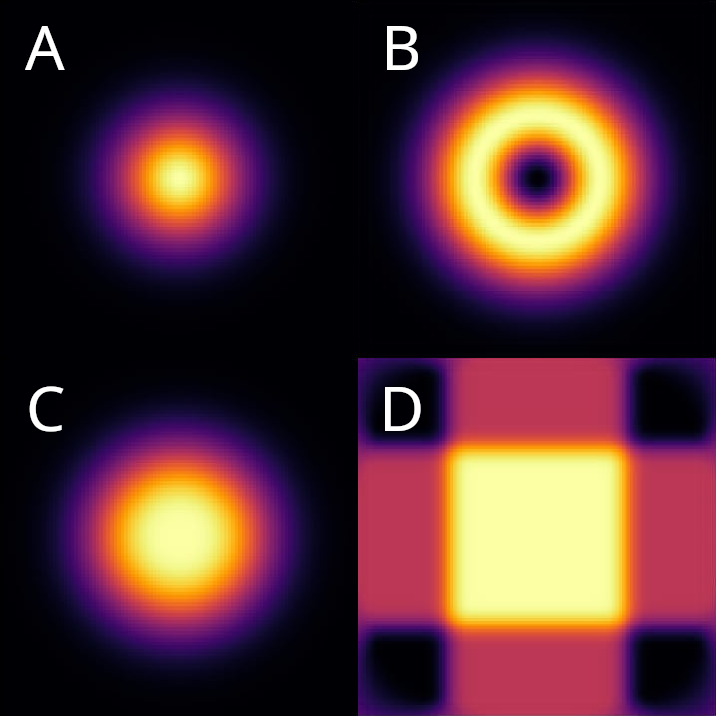}
    \caption{Visualization of primitives used for this work. The Husimi functions of (A) $\ket{0}_{\text{Fock}}$, (B) $\ket{1}_{\text{Fock}}$, and (C) $\left(\ketbra{0}_{\text{Fock}}+\ketbra{1}_{\text{Fock}}\right)/2$ are plotted. (D) is a square iteratively constructed as Eq.~\eqref{eq:map_square}.}
    \label{fig:primitives}
\end{figure}

For circles, we exploit the representation of Fock states in phase space instead of using a QIFS method or a modified version of the IFS method discussed in Section~\ref{sec:cfs}. As shown in panel~(A) in Fig.~\ref{fig:primitives}, the vacuum state $\ket{0}_{\text{Fock}}$ depicts a Gaussian at $(q,p)=(0,0)$ in phase space. Panel~(B) shows that $\ket{1}_{\text{Fock}}$ is represented as a ring that is slightly larger than the Gaussian. A equal mixture of these two states $\ketbra{0}_{\text{Fock}}$, $\ketbra{1}_{\text{Fock}}$ gives a circle as shown in panel~(C). By continually adding Fock states, a larger circle is obtained as long as the truncated system is large enough.
A equal mixture of all the Fock states in a (truncated) $d$-dimensional system, corresponds to a maximally mixed state, and the probability is uniformly distributed in phase space.
There are also certain states known as circular states, which are superpositions of coherent states~\cite{titulaer1966density,bialynicka1968properties,pathak2006generation} that are also pure states, unlike ours. If we would like to use more quantum-forward techniques with our method at some stage, the purity of states may become important and these states may become useful as well.



\subsubsection{Squares}
\label{sec:squares_quantum}

In this section, we will create a square iteratively with a QIFS method.
Consider a $d$-dimensional system with $d\times d$ pixels in the Husimi function. 
By converting Eqs.~\eqref{eqn:sq4} to their quantum counterparts and by following \cite{lozinski2003quantum}, a QIFS to create a square that expands on $L\times L$ pixels and is distant from the edge by $M$ pixels is given by
\begin{subequations} \label{eq:map_square}
\begin{align}
    \mathcal{G}_1\left[\rho\right]
    &\equiv
    \sum_{i,j=1}^{L}
    \ketbra{i+M}{j+M}_q
    \nonumber\\
    &\quad\quad\quad\quad
    \left(
    \sum_{n=1}^{d/L}
    \ketbra{(i-1)d/L+n}_q
    \right)
    \\
    \mathcal{G}_2\left[\rho\right]
    &\equiv
    \sum_{i,j=1}^{L}
    \ketbra{i+M+L}{j+M+L}_q
    \nonumber\\
    &\quad\quad\quad\quad
    \left(
    \sum_{n=1}^{d/L}
    \ketbra{(i-1)d/L+n}_q
    \right)
    \\
    \mathcal{G}_3\left[\rho\right]
    &\equiv
    \sum_{i,j=1}^{L}
    \ketbra{i+M}{j+M}_p
    \nonumber\\
    &\quad\quad\quad\quad
    \left(
    \sum_{n=1}^{d/L}
    \ketbra{(i-1)d/L+n}_p
    \right)
    \\
    \mathcal{G}_4\left[\rho\right]
    &\equiv
    \sum_{i,j=1}^{L}
    \ketbra{i+M+L}{j+M+L}_p
    \nonumber\\
    &\quad\quad\quad\quad
    \left(
    \sum_{n=1}^{d/L}
    \ketbra{(i-1)d/L+n}_p
    \right)
    ,
\end{align}
\end{subequations}
where $\ket{m}_q$ is a position basis and $\ket{m}_p$ is a momentum basis, assuming that $d/L$ is an integer.
The position and momentum operators $\hat{q},\hat{p}$ are given by
\begin{subequations}
\begin{align}
    \hat{q}
    &=
    \frac{1}{\sqrt{2}}
    \left(
    \hat{a}^{\dagger} + \hat{a}
    \right)
    \\
    \hat{p}
    &=
    \frac{i}{\sqrt{2}}
    \left(
    \hat{a}^{\dagger} - \hat{a}
    \right)
    ,
\end{align}
\end{subequations}
and their bases are the eigenstates of these operators. Also, these bases are related with Fourier transformation, i.e. 
\begin{align} \label{eq:fourier}
  \ket{k}_p
  =
  \frac{1}{\sqrt{N}}
  \sum_{j=1}^{N}e^{-i2\pi jk/N}\ket{j}_q  
  .
\end{align}

An example is shown in Fig.~\ref{fig:primitives}(D). For this method, non-zero probability is seen outside of the square that can be ignored by setting the threshold and extract the shape of a square. We will work on the creation of cleaner square in the future.

\subsection{Affine maps}
\label{sec:affine}

At this stage, we have shown the generation of two object primitives as well Gaussian and ring states that may also act as primitives.
Now it is important to focus on necessary transformations to allow for general-purpose rendering.
Affine maps might be the most common type of transformation expected by users and are typically implemented as an augmented matrix multiply like so:
\begin{equation}
    \begin{pmatrix}
        a && b && e \\
        c && d && f \\
        0 && 0 && 1 \\
    \end{pmatrix}
    \begin{pmatrix}
        P.x \\
        P.y \\
        1
    \end{pmatrix}
    \label{eqn:affine1}
\end{equation}
This formalism allows for rotations and shears with four variables in the upper-left ($a$, $b$, $c$, and $d$) representing scaling, rotation, and shearing operations, and the two variables on the right ($e$ and $f$) representing translation.
The final row composed of $(0, 0, 1 | 1)$ represents a hypothetical $z$ component that is unused in two-dimensional maps.
Another common formalism is the expansion of the matrix multiplication as:
\begin{subequations}
\label{eqn:affine2}
\begin{align}
    P.x &= aP.x + bP.y + e \\
    P.y &= cP.x + bP.y + f
\end{align}
\end{subequations}
To replicate this behavior on quantum architectures, we need methods to implement displacement, rotation, shearing, and scaling on quantum systems.
We will discuss each of these operations separately.

\subsubsection{Displacement}
The displacement operator in phase space has been well studied, e.g.~\cite{dodonov2002nonclassical}, and proposed by \cite{feynman1951operator,glauber1951some}. The author of the former is the founder of quantum simulation~\cite{feynman1982simulating}. It is defined as
\begin{equation}
    \hat{D}(\alpha)
    =
    e^{\alpha\hat{a}^{\dagger}-\alpha^*\hat{a}}
\end{equation}
with $\alpha$ a complex number.
Indeed, a coherent state can be written with a displacement operator for $\alpha=q+ip$ as
\begin{equation}
    \ket{q,p}
    =
    \hat{D}(\alpha)\ket{0}
    ,
\end{equation}
where $\ket{0}$ is a vacuum state. 
Following the representation of Eq.~\eqref{eqn:affine1}, the displacement operator updates the position and momentum operators $\hat{q},\hat{p}$
as follows:
\begin{align}
\begin{pmatrix}
    1 & 0 & \Re[\alpha]
    \\
    0 & 1 & \Im[\alpha]
    \\
    0 & 0 & 1
\end{pmatrix}
\begin{pmatrix}
    \hat{q} \\ \hat{p} \\ 1
\end{pmatrix}
.
\end{align}

\subsubsection{Rotation}

The rotation operator is defined as 
\begin{align}
    \hat{R}(\theta)
    &=
    e^{i\theta \hat{n}}
    ,
\end{align}
which adds a phase to Fock states.
Note that Fourier transformation~\eqref{eq:fourier} can be considered as a $\pi/2$ rotation~\cite{pei1997improved,candan2000discrete}.
The rotation operator rotates $\hat{q},\hat{p}$
as follows:
\begin{align}
\begin{pmatrix}
    \cos\theta & -\sin\theta & 0
    \\
    \sin\theta & \cos\theta & 0
    \\
    0 & 0 & 1
\end{pmatrix}
\begin{pmatrix}
    \hat{q} \\ \hat{p} \\ 1
\end{pmatrix}
.
\end{align}

\subsubsection{Squeezing}

The squeezing operator is a popular example for creating non-classical states~\cite{kitagawa1993squeezed,dodonov2002nonclassical}, and has attracted particular attention in metrology~\cite{pezze2018quantum}. 
The squeezing operator is defined as
\begin{equation}
    \hat{S}(z)
    =
    e^{\left(
    z\hat{a}^2 - z^*\hat{a}^{\dagger 2}
    \right)/2}
\end{equation}
with $z\equiv r e^{i\varphi}$ a complex number. As done in the above subsections, the squeezing operator updates as follows:
\begin{align}
\begin{pmatrix}
    \cosh{r}-\sinh{r}\cos{\varphi} 
    & 
    -\sinh{r}\sin{\varphi} & 0
    \\
    -\sinh{r}\sin{\varphi} 
    &
    \cosh{r}+\sinh{r}\cos{\varphi} & 0
    \\
    0 & 0 & 1
\end{pmatrix}
\begin{pmatrix}
    \hat{q} \\ \hat{p} \\ 1
\end{pmatrix}
.
\end{align}
As seen, $r$ is the strength and $\varphi$ is the angle of squeezing.
For convenience, we note the case of $\varphi=0$,  
\begin{align}
\begin{pmatrix}
    1/s 
    & 
    0 & 0
    \\
    0
    &
    s & 0
    \\
    0 & 0 & 1
\end{pmatrix}
\begin{pmatrix}
    \hat{q} \\ \hat{p} \\ 1
\end{pmatrix}
\end{align}
with $r=\log{s}$.

\subsubsection{Shearing}
The shearing operator is not as well known as the operators introduced above but, is important for general purpose computer graphics~\cite{gu2009quantum}. It is defined as 
\begin{equation}
    \hat{S}'(\beta)
    =
    e^{i\beta \hat{q}^2/2}
    ,
\end{equation}
and it updates $\hat{q},\hat{p}$ as follows: 
\begin{align}
\begin{pmatrix}
    1 & 0 & 0
    \\
    \beta & 1 & 0
    \\
    0 & 0 & 1
\end{pmatrix}
\begin{pmatrix}
    \hat{q} \\ \hat{p} \\ 1
\end{pmatrix}
.
\end{align}
Obviously, $e^{i\beta \hat{p}^2/2}$ is also a shearing operator and updates $\hat{q}\to\hat{q}+\beta\hat{p}$. 

\subsubsection{Expansion}

Expansion of a probability distribution leads to loss of information, and \cite{andreev2017scale} shows that expanding the Husimi function of a pure state results in a mixed state. For expanding mixed states, the same channel can be applied as any mixed state can be written as a set of pure states.
Thus, a map for expansion of $\rho$ is defined as
\begin{align}
    \label{eq:map_expansion}
    \mathcal{E}_{\lambda}[\rho]
    &=
    \sum_{j=0}^{\infty}
    \frac{\lambda^2\left(1-\lambda^2\right)^j}{j!}
    \sum_{k=0}^{\infty}
    \sqrt{\frac{(k+j)!}{k!}}
    \lambda^k
    \sum_{s=0}^{\infty}
    \sqrt{\frac{(s+j)!}{s!}}
    \lambda^s
    \nonumber\\
    &\quad\quad\quad\quad\quad\quad\quad\quad\quad\quad\quad
    \mel{k}{\rho}{s}_{\text{Fock}}
    \ketbra{k+j}{s+j}_{\text{Fock}}
\end{align}
with $\lambda<1$.
To be clear, the expansion is expressed as
\begin{align}
\begin{pmatrix}
    1/\lambda & 0 & 0
    \\
    0 & 1/\lambda & 0
    \\
    0 & 0 & 1
\end{pmatrix}
\begin{pmatrix}
    \hat{q} \\ \hat{p} \\ 1
\end{pmatrix}
.
\end{align}

\subsubsection{Contraction}

In contrast to expansion, it is difficult or perhaps impossible to perform contraction. For pure states, the expansion of the probability distribution is forbidden due to uncertainty principle. 
For mixed states, one may think about the reverse of the map for expansion~\eqref{eq:map_expansion}, but it is irreversible.
Nevertheless, it is easy to shrink the Husimi function after all other transformations, because it can be done by modifying the reference coherent state of the Husimi function or taking more qubits so that the system size increases.
Put simply, quantum primitives also have a primitive size that cannot be contracted, so we instead scale the output layer for contractive mappings.
This is discussed further in Section~\ref{sec:example}.






\subsection{General-purpose transformations}
\label{sec:gp}

Classical CFSs allow for manipulation of objects by non-affine transformations usually expressed as functional code-blocks.
The implementation in ~\cite{schloss2026cfs} provides a metaprogramming framework for this; however, as mentioned in Section~\ref{sec:related}, general-purpose programmatic workflows are not readily available on quantum devices at the current time.
For this reason, we will follow a similar construction as in Section~\ref{sec:affine} by showcasing some operations one might expect in quantum systems but limit ourselves to two simple use-cases: exponentiation and duplication.


\subsubsection{Exponentiation (Smears)}

Smear frames are essential in high-impact animation to express directionality in movement and emulate motion blur. They are also straightforward to implement with classical CFSs with non-affine transformations. 
In the quantum version, one can implement transformations, such as $\hat{p}\to\hat{p}+\gamma\hat{q}^k$, for positive integer $k$. 
Similar to shearing, such an operator is defined as 
\begin{equation} \label{eq:op_exponentiation}
    \hat{S}'_{k}(\gamma)
    =
    e^{i\gamma \hat{q}^{k+1}}
\end{equation}
with $\gamma$ real. This operation does not change $\hat{q}$ but updates $\hat{p}$ as $\hat{p}\to\hat{p}+(k+1)\gamma\hat{q}^{k}$. For $k=1$, it corresponds to shearing, and for $k\geq 2$ it is not an affine transformation anymore. Note that for $k=2$ it corresponds to cubic phase gate~\cite{gottesman2001encoding,gu2009quantum}, which is studied in the field of continuous variables.

Figure~\ref{fig:smiley} depicts smiley faces, and this operator $\hat{S}'_{k}(\gamma)$ is employed for the mouth.
Also, Fig.~\ref{fig:emuatom} describes electrons around an atom, and these operators $\hat{S}'_{k}(\gamma)$ are employed for specific transforms shown in Figure~\ref{fig:classical}(B).


\subsubsection{Duplication} \label{sec:duplication}

Duplicating objects is essential for constructing complex graphics. One way is to insert ancillary qubits and add control gates such that, depending on the configuration of the ancilla, the user can manipulate different objects. 
For instance, we may denote the state of a circle shown in Section~\ref{sec:circles_quantum} as $\rho_{\text{circle}}$, and we may express two circles with $\left(\ketbra{\downarrow}\otimes\rho_{\text{circle}}+\ketbra{\uparrow}\otimes\rho_{\text{circle}}\right)/2$ and also apply a control gate $\hat{D}(-5)\oplus\hat{D}(5)$ such that a circle moves left and the other moves right.

On the other hand, we have found that such control gates bring significant noise in actual computation on quantum hardware. Therefore, we have another implementation that uses a single Husimi function per object and combines them at a later stage in the rendering pipeline. Mathematically, this approach gives the same Husimi function as the ancilla approach except normalization, i.e. the Husimi function for a mixed state $\rho=\sum_{j} p_j \ketbra{\psi_j}$ is given by
\begin{equation}
    H_{\rho}(p,q)
    =
    \sum_{j} p_j
    H_{\ketbra{\psi_j}}(p,q)
    .
\end{equation}
This approach is to measure each of the Husimi function in the right-hand-side in the above equation one by one while the ancilla approach is to measure the term on the left-hand-side.
The downside is that, while the ancilla approach does not increase the render time for more objects, this approach increases the rendering time linearly. We note that this improvement by use of ancillas is not quantum advantage but simply due to the structure of a bit system.

\subsubsection{Example: smiley face}

\begin{figure*}
\includegraphics[width=0.8\textwidth]{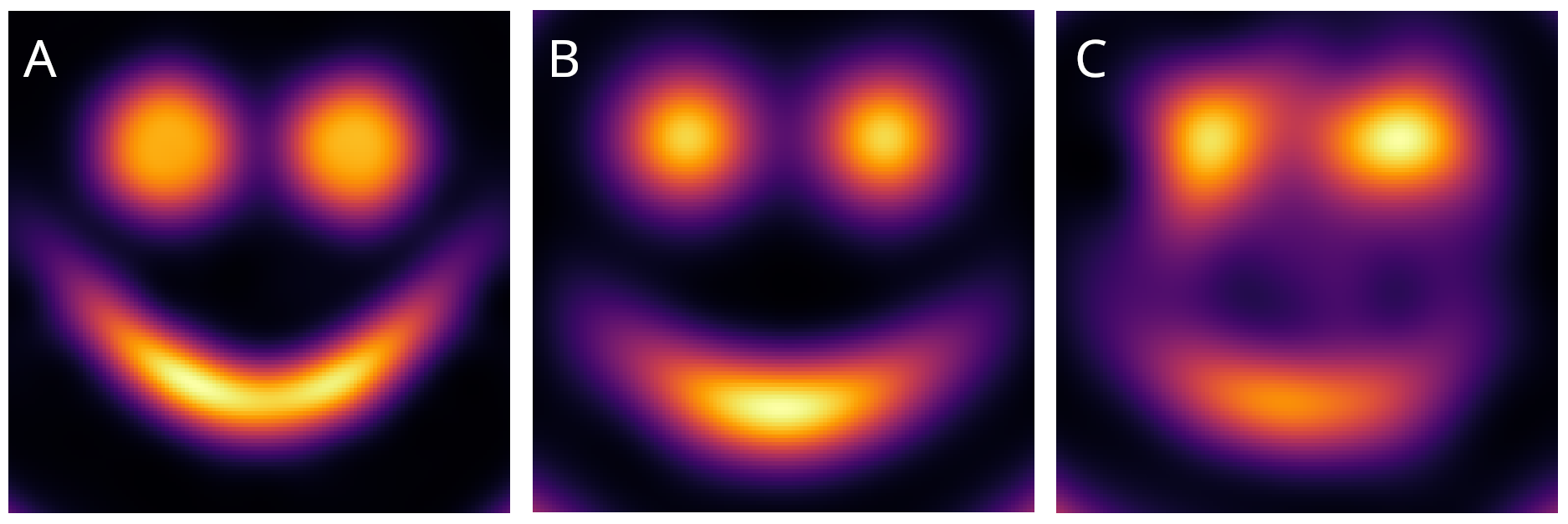}

\caption{visualization of a smiley face with (A) 5 qubits per mode and a circle primitive run on emulator, (B) 4 qubits per mode and a Gaussian primitive run on emulator, and (C) 4 qubits per mode and Gaussian primitive run on IBM's Kingston quantum computer. These images showcase duplication and smears, as described in the text.}
\label{fig:smiley}
\end{figure*}

Figure~\ref{fig:smiley} shows emulation of our approach on quantum hardware in the creation of a smiley face. The primitive for panel~(A) is a single circle, while panels~(B) and (C) use a Gaussian.
Panels~(A) and (B) use 2 ancillary qubits, but panel~(C) uses a single Husimi function per object.
An exponentiation operators for $k=2$ are applied to two primitives to create the mouth.
We will give more details of the implementation in the next section.





\begin{figure*}
\includegraphics[width=\textwidth]{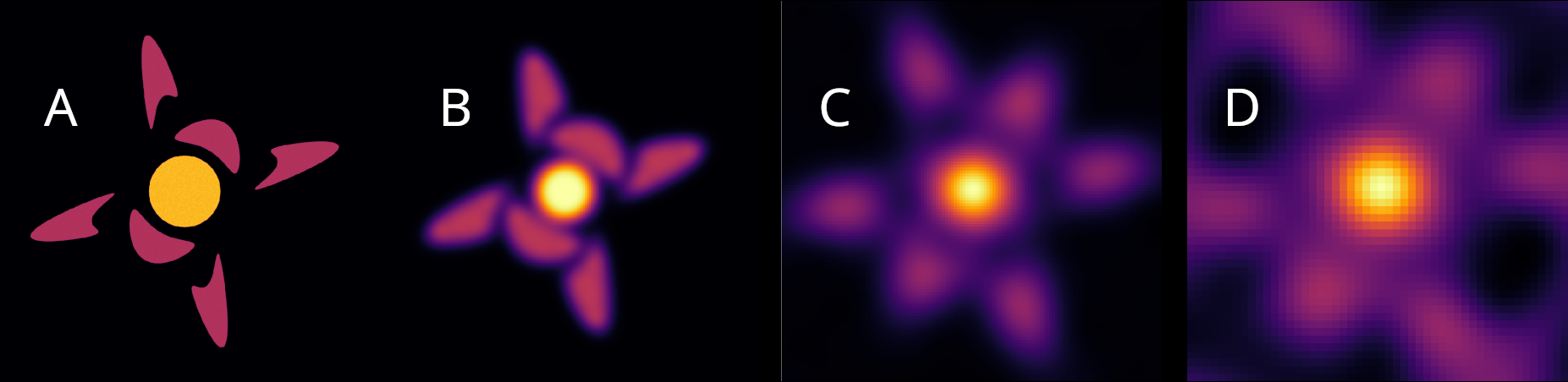}
\caption{
A comparison between the same frame from four different visualization techniques. (A) is a composable function system render, plotting the number of points in each pixel bin. (B) is a classical simulation of quantum architectures with 7 qubits per mode. (C) is an emulated version with noise expected from hardware (IBM’s Kingston quantum computer) and is restricted to 5 qubits per mode. (D) is a result on IBM’s Kingston quantum computer and is limited to 4 qubits per mode.
}
\label{fig:emuatom}
\end{figure*}

\subsection{An integrated example}
\label{sec:example}

\begin{figure}
\includegraphics[width=0.4\textwidth]{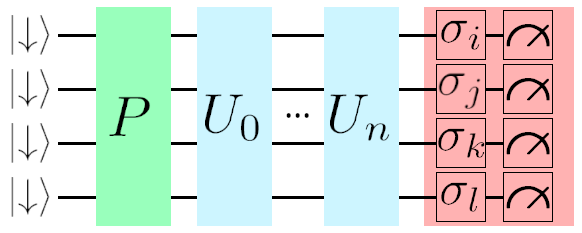}
\caption{
Sketch of the circuits for quantum CFSs. ``$P$'' represents a set of operators that make a primitive from the initial state $\ket{\downarrow\downarrow\downarrow\downarrow}$. ``$U_0,\ldots,U_n$'' represent a sequence of operators that perform affine and/or non-affine transformations. The left block performs state-tomography.
}
\label{fig:circuits}
\end{figure}

Now that we have seen the emulated results of all available transformations in Sections~\ref{sec:affine} and \ref{sec:gp}, we will now return to the complex visual in Section~\ref{sec:cfs} to replicate on quantum architectures.

First, we explain the sketch of the circuits (see Fig.~\ref{fig:circuits}). The initial state given in IBM hardware is $\ket{\downarrow\downarrow\ldots}$, and we regard it as a vacuum state $\ket{0}_{\text{Fock}}$. 
To create a primitive, we can make a circle by mixing Fock states as discussed in Section~\ref{sec:circles_quantum} for summing up the Husimi functions for different initial Fock states in the same way as explained in Section~\ref{sec:duplication}.
Due to the limited time available for running processes on quantum hardware, we have compromised on the number of Fock states to be prepared into only one, i.e. the primitive is a Gaussian for our hardware run. 
For affine and non-affine transformations, we program them as \texttt{UnitaryGate}s in Qiskit, except for the expansion channel~\eqref{eq:map_expansion}. Currently, Qiskit has classes for creating channels, such as Kraus operators and Super operators.
These are available to run with emulation, but do not have automate functions that make channels for hardware. Therefore, we have focused on unitary transformations for this study.
To construct Husimi function, we can directly measure the fidelity between the state of interest and a coherent state at each pixel on hardware or perform state tomography to extract the density matrix and then compute the Husimi function on a classical computer. The latter appears wasteful at first glance as it essentially performs tomography twice; however, considering the time required for preparing gates of coherent states and the fact that we need a large number of different circuits (one for each pixel), the latter is actually faster in practice.

Figures~\ref{fig:smiley}(B,C) show results of emulation and hardware run in the example of a smiley face, respectively. In contrast to panel~(A), we have used Gaussian as primitives instead of circles. 
The result on hardware suffer from noise from every operation and it is thus expected that all shapes are deformed to some extent. 
Nonetheless, we note that the probability distribution for the smile appears less distinct than for the eyes in (C).
This is because the smear operator~\eqref{eq:op_exponentiation} requires more two-qubit gates, which are more noisy in practice.

Figure~\ref{fig:emuatom} shows results of (A) classical CFSs, (B) classical simulation, (C) emulation with 5 qubits per mode, and (D) emulation with 4 qubits per mode in the example of visualization of an atom shown in Fig.~\ref{fig:classical}. 

In comparing the images, (A) and (B) appear to be strikingly similar, except for the Gaussian blur brought about by the Husimi function, as mentioned in Section~\ref{sec:husimi}; however, (C) and (D) are quite different from the others.
This is largely due to the number of qubits used to represent each mode.
Put simply, the more qubits used to represent the Fock state, the further away the coherent state can shift from the origin of the image without introducing nonphysical results.
Because the Husimi function output is dependent on the displacement operator, the world size ($W$) is dependent on the number of qubits used to represent each mode ($N$).
The rule of thumb is that your world size should be defined as $W=2^N$; however, different operations may shrink the usable space further.


This is why we are able to use the larger circle primitive for Figure~\ref{fig:smiley}(A), but must use the smaller Gaussian primitive for (B) and (C).
Though it is possible to represent a circle with 4 qubits, it is difficult to manipulate them with complex operations due to the limited world size, which is why we needed to change the operations to fit the usable space in Figure~\ref{fig:emuatom}(C) and (D).
Panel~(D), in particular, has so little usable space that the edges of the system are showing nonphysical noise.
Larger system sizes, like those shown in Figure~\ref{fig:smiley}(A) and Figure~\ref{fig:emuatom}(B) allow for more complex operations and primitives.
However, we have considered $N=4$ for two reasons:
\begin{enumerate}
\item For this work, we are interested in the minimal number of bits to represent complex scenes
\item Due to limitations in state tomography and using the Husimi function to output images, we both run out of memory when emulating larger systems and run out of time when using quantum architectures.
\end{enumerate}

There is no reason we could not draw the image of Figure~\ref{fig:emuatom}(B) on current-generation quantum hardware; however, it would take several minutes to render the frame.
Panel~(B), which uses 6 qubits, would take longer, and panel~(C), which uses 5 qubits, would take longer still.
Seeing this, it is now important to discuss the limitations to our method and how they may be resolved so we might be able to generate panel~(B) on current-generation hardware.

\subsection{Limitations and future work}
\label{sec:limits}

It is important to note that even though quantum systems may allow for several advantages over the classical implementation of this method, there are a still a number of limitations to this work that we are currently investigating and will leave to the future.

First, from a perspective of computer graphics we do not currently have a straightforward method to implement shading techniques, such as with fragment shaders.
As such, certain textures may be difficult to realize on quantum systems without solving the inverse problem akin to the difficulties in fractal compression~\cite{fisher1994fractal}.
However, seeing as how IFS and QIFS can be used as primitives for this work, this method might allow for study into such methods in the future.

There is also no straightforward programmatic interface that allows for sharing code between classical and quantum systems.
That is to say that users cannot simply type the transformations they would like to use and send the corresponding operations to a QPU.
This limitation is shared between all quantum frameworks, and was particularly apparent when mapping the results of this study to hardware due to our reliance on the \texttt{UnitaryGate} abstraction from Qiskit to create the final gate configurations, which may not be optimal and accrue unnecessarily high errors.
Specifically, this meant that Figure~\ref{fig:smiley}(C) required a different strategy to merge all the objects to a single layer.
While Figure~\ref{fig:smiley}~(A) and Figure~\ref{fig:smiley}(B) could use ancillary qubits, Figure~\ref{fig:smiley}~(C) required us to run three separate circuits, one for the left eye, right eye, and smile.
As this method is composed of simple algebraic manipulations, we believe that it can be used as a model for creating appropriate software abstractions for general-purpose uses in future work.

From the perspective of physics, we have not yet implemented certain operations like contractions and non-unitary map, i.e. quantum channels, on hardware.
Even if the system provided is closed, it is possible to make a non-unitary map by labeling a group of qubits as a reservoir which does not have to be larger than the rest. 
Developing an interface that enables the user to make non-unitary maps is not only convenient for our approach but also could make board impact on the usability of Qiskit.

Another limitation of this method is that we have currently only considered visual output with the Husimi function, which is relatively slow and cumbersome on quantum architectures.
As a specific example, when using the Husimi function on hardware, we would require a large number of (relatively) quick fidelity measurements on each pixel.
As a back-of-the-envelope calculation, if we wanted a 100$\times$100 image with the default number of shots in Qiskit (1024) per pixel, we would need 10,240,000 shots to construct the image.
The time per shot varies with different hardware, but assuming 100$\mu$s for each one, it would take approximately 3 hours to render a small image.
All that while ignoring other costs, such as enqueueing either one large circuit or 10,000 smaller ones.
For this reason, we relied heavily on Qiskit's \texttt{StateTomography} modules for this study; however, we are investigating methods to limit the number of fidelity measurements per pixel with quasi-Monte Carlo techniques.
Regardless, it is clear that high-definition imagery requires either a different image representation or a more efficient Husimi solver, some of which have been previously investigated~\cite{terraneo2005quantum}.
As mentioned in Section~\ref{sec:husimi}, the blurriness of the visual output from our visualizations is primarily due to out usage of the coherent state as the reference for the Husimi function and by modifying the reference state, we may also change the output.

By addressing the points mentioned in this section, it should be entirely possible to run high-quality and performant visualizations, eventually working our way from Figure~\ref{fig:emuatom}(D) back to (A).

\section{Conclusion}
\label{sec:conclusion}

In this paper, we have detailed the creation of a quantum-compatible Composable Function System (CFSs) and have demonstrated its utility by showcasing the generation of object primitives (Figure~\ref{fig:primitives}), the creation of the first video from a real quantum system (IBM's Kingston; Figure~\ref{fig:video} and supplementary video~\cite{suppvid}), and other complex visualizations Figures~\ref{fig:smiley} and~\ref{fig:emuatom}.
These results should allow for the generation of general-purpose graphics on quantum devices and have lasting implications for quantum metrology.
It is important to note that the classical formulation for CFSs is relatively new and, as this paper shows, is interoperable with quantum in many ways.
Therefore, advances in the classical formulation will also be useful for quantum.
We are hopeful that the results from this paper will act as a stepping stone into more interesting computer graphics applications on quantum architectures.

\section*{Data availability}

The code used for this work is freely available on github for both classical~\cite{quibbledocs} and quantum~\cite{qifs} visualizations.
The video output are all available in the supplementary video~\cite{suppvid}.

\begin{acks}

This work has been submitted to the \textit{NEDO Challenge, Quantum Computing ``Solve Social Issues !''} for the C-9 category \textit{Provision of New Rendering Environment}.
The work has been developed alongside the LeiosLabs community of passionate programmers on twitch, youtube, github, etc.
A.U. is financially supported by JSPS Overseas Research Fellowships and acknowledges financial support from Spanish MCIN (MCIN/AEI/10.13039/501100011033, contract No. PID2022-141283NB-I00 and No. PID2022-139099NB-I00) with the support of FEDER funds.

\end{acks}

\bibliographystyle{ACM-Reference-Format}
\bibliography{literature}

@String{Computing = "Computing" }

@String{Computer = "{IEEE} Computer" }

@String{Springer = "Springer-Verlag" }

@article{barnsley1985iterated,
  title={Iterated function systems and the global construction of fractals},
  author={Barnsley, Michael F and Demko, Stephen},
  journal={Proceedings of the Royal Society of London. A. Mathematical and Physical Sciences},
  volume={399},
  number={1817},
  pages={243--275},
  year={1985},
  publisher={The Royal Society London}
}

@article{ghosh2022iterated,
  title={Iterated function systems: A comprehensive survey},
  author={Ghosh, Ramen and Marecek, Jakub},
  journal={arXiv preprint arXiv:2211.14661},
  year={2022}
}

@article{draves2008fractal,
  title={The fractal flame algorithm},
  author={Draves, Scott and Reckase, Erik},
  journal={Citeseerx. Recuperado de http://citeseerx. ist. psu. edu/viewdoc/summary},
  year={2008},
  publisher={Citeseer}
}

@article{fisher1994fractal,
  title={Fractal image compression},
  author={Fisher, Yuval},
  journal={Fractals},
  volume={2},
  number={03},
  pages={347--361},
  year={1994},
  publisher={World Scientific}
}

@article{barnsley2011chaos,
  title={The chaos game on a general iterated function system},
  author={Barnsley, Michael F and Vince, Andrew},
  journal={Ergodic theory and dynamical systems},
  volume={31},
  number={4},
  pages={1073--1079},
  year={2011},
  publisher={Cambridge University Press}
}

@inproceedings{zhang2021judging,
  title={Judging a type by its pointer: optimizing GPU virtual functions},
  author={Zhang, Mengchi and Alawneh, Ahmad and Rogers, Timothy G},
  booktitle={Proceedings of the 26th ACM International Conference on Architectural Support for Programming Languages and Operating Systems},
  pages={241--254},
  year={2021}
}

@incollection{slusallek2005introduction,
  title={Introduction to real-time ray tracing},
  author={Slusallek, Philipp and Shirley, Peter and Mark, William and Stoll, Gordon and Wald, Ingo},
  booktitle={ACM SIGGRAPH 2005 Courses},
  pages={1--es},
  year={2005}
}

@misc{quibbledocs,
  author = {James Schloss},
  title = {Quibble Docs},
  howpublished = "\url{http://www.leioslabs.com/quibble/}",
  year = {2026}, 
}

@article{elliott2003functional,
  title={Functional images},
  author={Elliott, Conal},
  journal={The Fun of Programming,“Cornerstones of Computing” series. Palgrave},
  year={2003}
}

@article{kuth2024real,
  title={Real-time procedural generation with GPU work graphs},
  author={Kuth, Bastian and Oberberger, Max and Faber, Carsten and Baumeister, Dominik and Chajdas, Matth{\"a}us and Meyer, Quirin},
  journal={Proceedings of the ACM on Computer Graphics and Interactive Techniques},
  volume={7},
  number={3},
  pages={1--16},
  year={2024},
  publisher={ACM New York, NY, USA}
}

@article{keeter2020massively,
  title={Massively parallel rendering of complex closed-form implicit surfaces},
  author={Keeter, Matthew J},
  journal={ACM Transactions on Graphics (TOG)},
  volume={39},
  number={4},
  pages={141--1},
  year={2020},
  publisher={ACM New York, NY, USA}
}

@misc{schloss2026cfs,
      title={Composable function systems as a general-purpose rendering framework}, 
      author={James Schloss},
      year={2026},
      eprint={2606.02226},
      archivePrefix={arXiv},
      primaryClass={cs.GR},
      url={https://arxiv.org/abs/2606.02226}, 
}

@article{zhu2026qllvm,
  title={QLLVM: A Scalable Quantum-Classical Co-Compilation Framework based on LLVM},
  author={Zhu, Yu and Du, Qiming and Jin, Yuqiong and He, Woji and Lian, Hang and Zhou, Xin and Xu, Jinchen and Shan, Zheng},
  journal={arXiv preprint arXiv:2604.15094},
  year={2026}
}

@article{vazquez2024qpu,
  title={QPU integration in OpenCL for heterogeneous programming: J. V{\'a}zquez-P{\'e}rez et al.},
  author={V{\'a}zquez-P{\'e}rez, Jorge and Pi{\~n}eiro, C{\'e}sar and Pichel, Juan C and Pena, Tom{\'a}s F and G{\'o}mez, Andr{\'e}s},
  journal={The Journal of Supercomputing},
  volume={80},
  number={8},
  pages={11682--11703},
  year={2024},
  publisher={Springer}
}

@article{javadi2024quantum,
  title={Quantum computing with Qiskit},
  author={Javadi-Abhari, Ali and Treinish, Matthew and Krsulich, Kevin and Wood, Christopher J and Lishman, Jake and Gacon, Julien and Martiel, Simon and Nation, Paul D and Bishop, Lev S and Cross, Andrew W and others},
  journal={arXiv preprint arXiv:2405.08810},
  year={2024}
}

@article{bergholm2018pennylane,
  title={Pennylane: Automatic differentiation of hybrid quantum-classical computations},
  author={Bergholm, Ville and Izaac, Josh and Schuld, Maria and Gogolin, Christian and Ahmed, Shahnawaz and Ajith, Vishnu and Alam, M Sohaib and Alonso-Linaje, Guillermo and AkashNarayanan, Bharath and Asadi, Ali and others},
  journal={arXiv preprint arXiv:1811.04968},
  year={2018}
}

@inproceedings{hopf2026integrating,
  title={Integrating Quantum Software Tools with (in) MLIR},
  author={Hopf, Patrick and Ochoa, Erick and Stade, Yannick and Rovara, Damian and Quetschlich, Nils and Florea, Ioan Albert and Izaac, Josh and Wille, Robert and Burgholzer, Lukas},
  booktitle={Proceedings of the Supercomputing Asia and International Conference on High Performance Computing in Asia Pacific Region},
  pages={42--54},
  year={2026}
}

@article{ma2011quantum,
  title={Quantum spin squeezing},
  author={Ma, Jian and Wang, Xiaoguang and Sun, Chang-Pu and Nori, Franco},
  journal={Physics Reports},
  volume={509},
  number={2-3},
  pages={89--165},
  year={2011},
  publisher={Elsevier}
}

@article{lozinski2003quantum,
  title={Quantum iterated function systems},
  author={{\L}ozi{\'n}ski, Artur and {\.Z}yczkowski, Karol and S{\l}omczy{\'n}ski, Wojciech},
  journal={Physical Review E},
  volume={68},
  number={4},
  pages={046110},
  year={2003},
  publisher={APS}
}

@article{jadczyk2004quantum,
  title={On quantum iterated function systems},
  author={Jadczyk, Arkadiusz},
  journal={Central European Journal of Physics},
  volume={2},
  number={3},
  pages={492--503},
  year={2004},
  publisher={Springer}
}

@article{scott2003bakers,
  title={Entangling power of the quantum baker's map},
  author={Scott, Andrew J and Caves, Carlton M},
  journal={Journal of Physics A: Mathematical and General},
  volume={36},
  number={36},
  pages={9553--9576},
  year={2003}
}

@article{balazs1987bakers,
  title={The quantized Baker's transformation},
  author={Balazs, NL and Voros, A},
  journal={EPL (Europhysics Letters)},
  volume={4},
  number={10},
  pages={1089--1094},
  year={1987}
}

@article{pakonski1999bakers,
  title={Quantum baker map on the sphere},
  author={Pakonski, Prot and Ostruszka, Andrzej and Zyczkowski, Karol},
  journal={Nonlinearity},
  volume={12},
  number={2},
  pages={269--284},
  year={1999}
}

@article{montenegro2025metrology,
  title={Quantum metrology and sensing with many-body systems},
  author={Montenegro, Victor and Mukhopadhyay, Chiranjib and Yousefjani, Rozhin and Sarkar, Saubhik and Mishra, Utkarsh and Paris, Matteo GA and Bayat, Abolfazl},
  journal={Physics Reports},
  volume={1134},
  pages={1--62},
  year={2025},
  publisher={Elsevier}
}

@article{schloss2020controlled,
  title={Controlled creation of three-dimensional vortex structures in Bose-Einstein condensates using artificial magnetic fields},
  author={Schloss, James and Barnett, Peter and Sachdeva, Rashi and Busch, Thomas},
  journal={Physical review a},
  volume={102},
  number={4},
  pages={043325},
  year={2020},
  publisher={APS}
}

@article{siegl2022controlled,
  title={Controlled creation of quantum skyrmions},
  author={Siegl, Pia and Vedmedenko, Elena Y and Stier, Martin and Thorwart, Michael and Posske, Thore},
  journal={Physical Review Research},
  volume={4},
  number={2},
  pages={023111},
  year={2022},
  publisher={APS}
}

@article{weinacht1999controlling,
  title={Controlling the shape of a quantum wavefunction},
  author={Weinacht, TC and Ahn, Jaewook and Bucksbaum, Phil H},
  journal={Nature},
  volume={397},
  number={6716},
  pages={233--235},
  year={1999},
  publisher={Nature Publishing Group UK London}
}

@article{bao2024creating,
  title={Creating and controlling global Greenberger-Horne-Zeilinger entanglement on quantum processors},
  author={Bao, Zehang and Xu, Shibo and Song, Zixuan and Wang, Ke and Xiang, Liang and Zhu, Zitian and Chen, Jiachen and Jin, Feitong and Zhu, Xuhao and Gao, Yu and others},
  journal={Nature Communications},
  volume={15},
  number={1},
  pages={8823},
  year={2024},
  publisher={Nature Publishing Group UK London}
}

@article{wang2022imagereview,
  title={Review of quantum image processing},
  author={Wang, Zhaobin and Xu, Minzhe and Zhang, Yaonan},
  journal={Archives of Computational Methods in Engineering},
  volume={29},
  number={2},
  pages={737--761},
  year={2022},
  publisher={Springer}
}

@article{deb2024imagecompress,
  title={Quantum image compression: Fundamentals, algorithms, and advances},
  author={Deb, Sowmik Kanti and Pan, W David},
  journal={Computers},
  volume={13},
  number={8},
  pages={185},
  year={2024},
  publisher={MDPI}
}

@article{roncallo2023jpeg,
  title={Quantum jpeg},
  author={Roncallo, Simone and Maccone, Lorenzo and Macchiavello, Chiara},
  journal={AVS Quantum Science},
  volume={5},
  number={4},
  year={2023},
  publisher={AIP Publishing}
}

@article{latorre2005image,
  title={Image compression and entanglement},
  author={Latorre, Jose I},
  journal={arXiv preprint quant-ph/0510031},
  year={2005}
}

@article{dhar2024watermark,
  title={Digital to quantum watermarking: A journey from past to present and into the future},
  author={Dhar, Swapnaneel and Sahu, Aditya Kumar},
  journal={Computer Science Review},
  volume={54},
  pages={100679},
  year={2024},
  publisher={Elsevier}
}

@article{husimi1940,
  title={Some formal properties of the density matrix},
  author={Husimi, K{\^o}di},
  journal={Proceedings of the Physico-Mathematical Society of Japan. 3rd Series},
  volume={22},
  number={4},
  pages={264--314},
  year={1940},
  publisher={The Physical Society of Japan, The Mathematical Society of Japan}
}

@article{wigner1932quantum,
  title={On the quantum correction for thermodynamic equilibrium},
  author={Wigner, Eugene},
  journal={Physical review},
  volume={40},
  number={5},
  pages={749},
  year={1932},
  publisher={APS}
}

@inproceedings{stavenger2022c2qa,
  title={C2qa-bosonic qiskit},
  author={Stavenger, Timothy J and Crane, Eleanor and Smith, Kevin C and Kang, Christopher T and Girvin, Steven M and Wiebe, Nathan},
  booktitle={2022 IEEE High Performance Extreme Computing Conference (HPEC)},
  pages={1--8},
  year={2022},
  organization={IEEE}
}

@article{kitagawa1993squeezed,
  title = {Squeezed spin states},
  author = {Kitagawa, Masahiro and Ueda, Masahito},
  journal = {Phys. Rev. A},
  volume = {47},
  issue = {6},
  pages = {5138--5143},
  numpages = {0},
  year = {1993},
  month = {Jun},
  publisher = {American Physical Society},
  doi = {10.1103/PhysRevA.47.5138},
  url = {https://link.aps.org/doi/10.1103/PhysRevA.47.5138}
}

@article{pezze2018quantum,
  title = {Quantum metrology with nonclassical states of atomic ensembles},
  author = {Pezz\`e, Luca and Smerzi, Augusto and Oberthaler, Markus K. and Schmied, Roman and Treutlein, Philipp},
  journal = {Rev. Mod. Phys.},
  volume = {90},
  issue = {3},
  pages = {035005},
  numpages = {70},
  year = {2018},
  month = {Sep},
  publisher = {American Physical Society},
  doi = {10.1103/RevModPhys.90.035005},
  url = {https://link.aps.org/doi/10.1103/RevModPhys.90.035005}
}

@article{lozinski2002irreversible,
  title = {Irreversible quantum baker map},
  author = {\L{}ozi\ifmmode \acute{n}\else \'{n}\fi{}ski, Artur and Pako\ifmmode \acute{n}\else \'{n}\fi{}ski, Prot and \ifmmode \dot{Z}\else \.{Z}\fi{}yczkowski, Karol},
  journal = {Phys. Rev. E},
  volume = {66},
  issue = {6},
  pages = {065201(R)},
  numpages = {4},
  year = {2002},
  month = {Dec},
  publisher = {American Physical Society},
  doi = {10.1103/PhysRevE.66.065201},
  url = {https://link.aps.org/doi/10.1103/PhysRevE.66.065201}
}

@article{balazs1989the,
title = {The quantized Baker's transformation},
journal = {Annals of Physics},
volume = {190},
number = {1},
pages = {1-31},
year = {1989},
issn = {0003-4916},
doi = {https://doi.org/10.1016/0003-4916(89)90259-5},
url = {https://www.sciencedirect.com/science/article/pii/0003491689902595},
author = {N.L. Balazs and A. Voros}
}

@article{dodonov2002nonclassical,
doi = {10.1088/1464-4266/4/1/201},
url = {https://doi.org/10.1088/1464-4266/4/1/201},
year = {2002},
month = {jan},
publisher = {},
volume = {4},
number = {1},
pages = {R1},
author = {V V Dodonov},
title = {`Nonclassical'
states in quantum optics: a `squeezed' review of the first 75 years},
journal = {Journal of Optics B: Quantum and Semiclassical Optics}
}

@article{feynman1951operator,
  title = {An Operator Calculus Having Applications in Quantum Electrodynamics},
  author = {Feynman, Richard P.},
  journal = {Phys. Rev.},
  volume = {84},
  issue = {1},
  pages = {108--128},
  numpages = {0},
  year = {1951},
  month = {Oct},
  publisher = {American Physical Society},
  doi = {10.1103/PhysRev.84.108},
  url = {https://link.aps.org/doi/10.1103/PhysRev.84.108}
}

@article{glauber1951some,
  title = {Some Notes on Multiple-Boson Processes},
  author = {Glauber, Roy J.},
  journal = {Phys. Rev.},
  volume = {84},
  issue = {3},
  pages = {395--400},
  numpages = {0},
  year = {1951},
  month = {Nov},
  publisher = {American Physical Society},
  doi = {10.1103/PhysRev.84.395},
  url = {https://link.aps.org/doi/10.1103/PhysRev.84.395}
}

@article{feynman1982simulating,
	author = {Feynman, Richard P. },
	date = {1982/06/01},
	doi = {10.1007/BF02650179},
	id = {Feynman1982},
	isbn = {1572-9575},
	journal = {International Journal of Theoretical Physics},
	number = {6},
	pages = {467--488},
	title = {Simulating physics with computers},
	url = {https://doi.org/10.1007/BF02650179},
	volume = {21},
	year = {1982}}

@ARTICLE{candan2000discrete,
  author={Candan, C. and Kutay, M.A. and Ozaktas, H.M.},
  journal={IEEE Transactions on Signal Processing}, 
  title={The discrete fractional Fourier transform}, 
  year={2000},
  volume={48},
  number={5},
  pages={1329-1337},
  doi={10.1109/78.839980}}

@article{pei1997improved,
author = {Soo-Chang Pei and Min-Hung Yeh},
journal = {Opt. Lett.},
number = {14},
pages = {1047--1049},
publisher = {Optica Publishing Group},
title = {Improved discrete fractional Fourier transform},
volume = {22},
month = {Jul},
year = {1997},
url = {https://opg.optica.org/ol/abstract.cfm?URI=ol-22-14-1047},
doi = {10.1364/OL.22.001047}
}

@article{gottesman2001encoding,
  title = {Encoding a qubit in an oscillator},
  author = {Gottesman, Daniel and Kitaev, Alexei and Preskill, John},
  journal = {Phys. Rev. A},
  volume = {64},
  issue = {1},
  pages = {012310},
  numpages = {21},
  year = {2001},
  month = {Jun},
  publisher = {American Physical Society},
  doi = {10.1103/PhysRevA.64.012310},
  url = {https://link.aps.org/doi/10.1103/PhysRevA.64.012310}
}

@article{madsen2022quantum,
	author = {Madsen, Lars S. and Laudenbach, Fabian and Askarani, Mohsen Falamarzi. and Rortais, Fabien and Vincent, Trevor and Bulmer, Jacob F. F. and Miatto, Filippo M. and Neuhaus, Leonhard and Helt, Lukas G. and Collins, Matthew J. and Lita, Adriana E. and Gerrits, Thomas and Nam, Sae Woo and Vaidya, Varun D. and Menotti, Matteo and Dhand, Ish and Vernon, Zachary and Quesada, Nicol{\'a}s and Lavoie, Jonathan},
	date = {2022/06/01},
	doi = {10.1038/s41586-022-04725-x},
	id = {Madsen2022},
	isbn = {1476-4687},
	journal = {Nature},
	number = {7912},
	pages = {75--81},
	title = {Quantum computational advantage with a programmable photonic processor},
	url = {https://doi.org/10.1038/s41586-022-04725-x},
	volume = {606},
	year = {2022}}

@article{aghaee2025scaling,
	author = {Aghaee Rad, H. and Ainsworth, T. and Alexander, R. N. and Altieri, B. and Askarani, M. F. and Baby, R. and Banchi, L. and Baragiola, B. Q. and Bourassa, J. E. and Chadwick, R. S. and Charania, I. and Chen, H. and Collins, M. J. and Contu, P. and D'Arcy, N. and Dauphinais, G. and De Prins, R. and Deschenes, D. and Di Luch, I. and Duque, S. and Edke, P. and Fayer, S. E. and Ferracin, S. and Ferretti, H. and Gefaell, J. and Glancy, S. and Gonz{\'a}lez-Arciniegas, C. and Grainge, T. and Han, Z. and Hastrup, J. and Helt, L. G. and Hillmann, T. and Hundal, J. and Izumi, S. and Jaeken, T. and Jonas, M. and Kocsis, S. and Krasnokutska, I. and Larsen, M. V. and Laskowski, P. and Laudenbach, F. and Lavoie, J. and Li, M. and Lomonte, E. and Lopetegui, C. E. and Luey, B. and Lund, A. P. and Ma, C. and Madsen, L. S. and Mahler, D. H. and Mantilla Calder{\'o}n, L. and Menotti, M. and Miatto, F. M. and Morrison, B. and Nadkarni, P. J. and Nakamura, T. and Neuhaus, L. and Niu, Z. and Noro, R. and Papirov, K. and Pesah, A. and Phillips, D. S. and Plick, W. N. and Rogalsky, T. and Rortais, F. and Sabines-Chesterking, J. and Safavi-Bayat, S. and Sazhaev, E. and Seymour, M. and Rezaei Shad, K. and Silverman, M. and Srinivasan, S. A. and Stephan, M. and Tang, Q. Y. and Tasker, J. F. and Teo, Y. S. and Then, R. B. and Tremblay, J. E. and Tzitrin, I. and Vaidya, V. D. and Vasmer, M. and Vernon, Z. and Villalobos, L. F. S. S. M. and Walshe, B. W. and Weil, R. and Xin, X. and Yan, X. and Yao, Y. and Zamani Abnili, M. and Zhang, Y.},
	date = {2025/02/01},
	doi = {10.1038/s41586-024-08406-9},
	id = {Aghaee Rad2025},
	isbn = {1476-4687},
	journal = {Nature},
	number = {8052},
	pages = {912--919},
	title = {Scaling and networking a modular photonic quantum computer},
	url = {https://doi.org/10.1038/s41586-024-08406-9},
	volume = {638},
	year = {2025}}

@article{gu2009quantum,
  title = {Quantum computing with continuous-variable clusters},
  author = {Gu, Mile and Weedbrook, Christian and Menicucci, Nicolas C. and Ralph, Timothy C. and van Loock, Peter},
  journal = {Phys. Rev. A},
  volume = {79},
  issue = {6},
  pages = {062318},
  numpages = {16},
  year = {2009},
  month = {Jun},
  publisher = {American Physical Society},
  doi = {10.1103/PhysRevA.79.062318},
  url = {https://link.aps.org/doi/10.1103/PhysRevA.79.062318}
}

@article{andreev2017scale,
	author = {Andreev, V. A. and Davidovi{\'c}, D. M. and Davidovi{\'c}, L. D. and Davidovi{\'c}, Milena D. and Davidovi{\'c}, Milo{\v s}D.},
	date = {2017/07/01},
	doi = {10.1134/S0040577917070091},
	id = {Andreev2017},
	isbn = {1573-9333},
	journal = {Theoretical and Mathematical Physics},
	number = {1},
	pages = {1080--1096},
	title = {Scale transformations in phase space and stretched states of a harmonic oscillator},
	url = {https://doi.org/10.1134/S0040577917070091},
	volume = {192},
	year = {2017}}

@article{lesniak2025highly,
  title={Highly non-contractive iterated function systems on Euclidean space can have an attractor},
  author={Le{\'s}niak, Krzysztof and Snigireva, Nina and Strobin, Filip and Vince, Andrew},
  journal={Journal of Dynamics and Differential Equations},
  volume={37},
  number={3},
  pages={2371--2388},
  year={2025},
  publisher={Springer}
}

@misc{qifs,
  author = {Schloss, James and Usui, Ayaka},
  title = {{QIFS.jl}},
  howpublished = "\url{https://github.com/leios/QIFS.jl}",
  year = {2026}, 
  note = "work in progress"
}

@misc{suppvid,
  author = {Schloss, James and Usui, Ayaka},
  title = {{Supplementary Video}},
  howpublished = "\url{https://youtu.be/2fdknkNNb4s}",
  year = {2026}, 
  note = "To be added to supplementary material upon publication"
}

@article{terraneo2005quantum,
  title = {Quantum computation and analysis of Wigner and Husimi functions: Toward a quantum image treatment},
  author = {Terraneo, M. and Georgeot, B. and Shepelyansky, D. L.},
  journal = {Phys. Rev. E},
  volume = {71},
  issue = {6},
  pages = {066215},
  numpages = {14},
  year = {2005},
  month = {Jun},
  publisher = {American Physical Society},
  doi = {10.1103/PhysRevE.71.066215},
  url = {https://link.aps.org/doi/10.1103/PhysRevE.71.066215}
}

@article{titulaer1966density,
  title = {Density Operators for Coherent Fields},
  author = {Titulaer, U. M. and Glauber, R. J.},
  journal = {Phys. Rev.},
  volume = {145},
  issue = {4},
  pages = {1041--1050},
  numpages = {0},
  year = {1966},
  month = {May},
  publisher = {American Physical Society},
  doi = {10.1103/PhysRev.145.1041},
  url = {https://link.aps.org/doi/10.1103/PhysRev.145.1041}
}

@article{bialynicka1968properties,
  title = {Properties of the Generalized Coherent State},
  author = {Bialynicka---Birula, Z.},
  journal = {Phys. Rev.},
  volume = {173},
  issue = {5},
  pages = {1207--1209},
  numpages = {0},
  year = {1968},
  month = {Sep},
  publisher = {American Physical Society},
  doi = {10.1103/PhysRev.173.1207},
  url = {https://link.aps.org/doi/10.1103/PhysRev.173.1207}
}

@article{pathak2006generation,
  title = {Generation of a superposition of multiple mesoscopic states of radiation in a resonant cavity},
  author = {Pathak, P. K. and Agarwal, G. S.},
  journal = {Phys. Rev. A},
  volume = {71},
  issue = {4},
  pages = {043823},
  numpages = {6},
  year = {2005},
  month = {Apr},
  publisher = {American Physical Society},
  doi = {10.1103/PhysRevA.71.043823},
  url = {https://link.aps.org/doi/10.1103/PhysRevA.71.043823}
}

\appendix

\section{Notes on classical Composable Function Systems}
\label{app:classysquare}
Though we have focused primarily on circle and Gaussian primitives for this work, other primitives, like the square (or even the tartar map) can be used on quantum.
Moreover, we will use this section to highlight some interesting features of CFS workflows.
\begin{figure}
\includegraphics[width=0.4\textwidth]{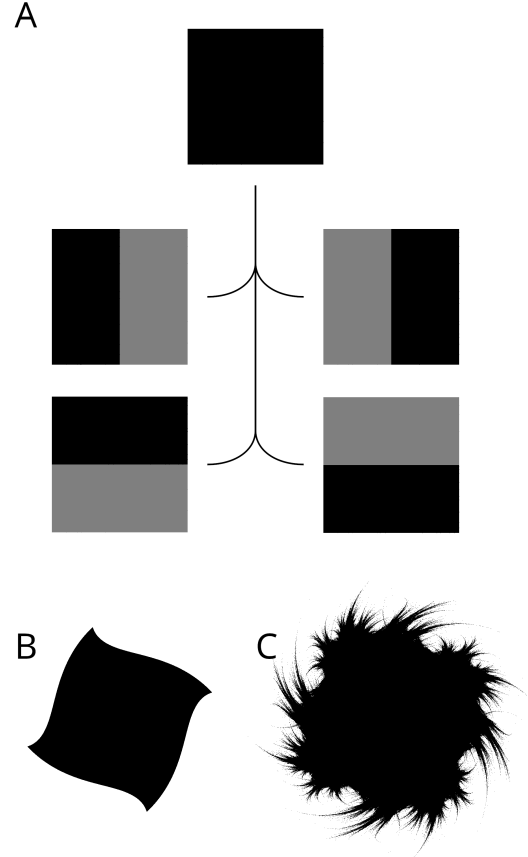}
\caption{IFS for the unit square with decoupled $x$ and $y$ components. (A) represents four distinct transformations ($f_n \in {1,2,3,4}$) as described in the text (Equations~\ref{eqn:sq1}-\ref{eqn:sq4}). (B) is an arbitrary transformation on the square.}
\label{fig:square}
\end{figure}

First, let us consider the following IFS for a square:

\begin{subequations}
\label{eqn:sq4}
\begin{align}
    f_1(P) &= \frac{P_x}{2} \label{eqn:sq1}\\
    f_2(P) &= \frac{P_x + 1}{2} \label{eqn:sq2}\\
    f_3(P) &= \frac{P_y}{2} \label{eqn:sq3}\\
    f_3(P) &= \frac{P_y + 1}{2}
\end{align}
\end{subequations}
Here, $P$ is a two dimensional point with an $x$ and $y$ component (labeled as $P_x$ and $P_y$), and the functions $f_n$ represent mappings of the square onto half of itself.
Concretely, $f_1$ maps the square to the left half of itself, $f_2$ to the right, $f_3$ to the bottom, and $f_4$ to the top.
These are pictorially shown in Figure~\ref{fig:square}(A).
It should be noted that there are many methods to generate a square with IFSs~\cite{balazs1987bakers,balazs1989the,schloss2026cfs}; however, this one allows for the construction of a square without coupling the $x$ and $y$ components.
That is to say that $f_1$ and $f_2$, only require $P.x$ and $f_3$ and $f_4$ only require $P.y$.
This IFS was chosen as it allows for more straightforward construction on QPUs.

Now, consider the following transformation in $\mathbb{R}^2$: 
\begin{align}
    r &= \sqrt{x^2 + y^2} \\
    x &= (x\cos(r^2) + y\sin(r^2)\\
    y &= (x\sin(r^2) + y\cos(r^2)\\
    P_2 &= (x,y)
\end{align}
Combining this equation with the IFS for the unit square (Equations~\ref{eqn:sq1}-\ref{eqn:sq4}) can result in either Figure~\ref{fig:square}(B) or (C) depending on how the functions are composed.
(B) is the result if a separate, copy point is used for $g$, while (C) is if the same point is used for both functions.
(B) is more likely what the user would expect and is similar to ~\cite{elliott2003functional}, but (C) is more closely related to methods like fractal flames~\cite{draves2008fractal}.
It is possible to chain functions together to precisely place points where the user desires and is therefore possible to create topologically non-trivial transformations, duplications, smears, etc.


\end{document}